\documentclass[conference,10pt]{IEEEtran}
\IEEEoverridecommandlockouts
\usepackage{cite}
\usepackage{amsmath,amssymb,amsfonts}
\usepackage{algorithmic}
\usepackage{graphicx}
\usepackage{subcaption}
\usepackage{textcomp}
\usepackage{xcolor}
\usepackage{url}
\usepackage{makecell}
\usepackage[export]{adjustbox}
\def\BibTeX{{\rm B\kern-.05em{\sc i\kern-.025em b}\kern-.08em
    T\kern-.1667em\lower.7ex\hbox{E}\kern-.125emX}}
    
\usepackage[bottom=0.73in,top=0.7in,left=0.57in,right=0.57in]{geometry}
\begin{document}


\title{Statistical Analysis of Received Signal Strength in Industrial IoT Distributed Massive MIMO Systems}

\author{Eduardo Noboro Tominaga\IEEEauthorrefmark{1}, Onel Luiz Alcaraz López\IEEEauthorrefmark{1}, Richard Demo Souza\IEEEauthorrefmark{2}, Hirley Alves\IEEEauthorrefmark{1},\\

	\IEEEauthorblockA{
		\IEEEauthorrefmark{1}6G Flagship, Centre for Wireless Communications (CWC), University of Oulu, Finland\\
		\{eduardo.noborotominaga, onel.alcarazlopez, hirley.alves\}@oulu.fi\\
		\IEEEauthorrefmark{2}Federal University of Santa Catarina (UFSC), Florian\'{o}polis, Brazil, richard.demo@ufsc.br\\
	}
}

\maketitle

\begin{abstract}
The Fifth Generation (5G) of wireless networks introduced native support for Machine-Type Communication (MTC), which is a key enabler for the Internet of Things (IoT) revolution. 
Current 5G standards are not yet capable of fully satisfying the requirements of critical MTC (cMTC) and massive MTC (mMTC) use cases. This is the main reason why industry and academia have already started working on technical solutions for beyond-5G and Sixth Generation (6G) networks. One technological solution that has been extensively studied is the combination of network densification, massive Multiple-Input Multiple-Output (mMIMO) systems and user-centric design, which is known as distributed mMIMO or Cell-Free (CF) mMIMO. Under this new paradigm, there are no longer cell boundaries: all the Access Points (APs) on the network cooperate to jointly serve all the devices. In this paper, we compare the performance of traditional mMIMO and different distributed mMIMO setups, and quantify the macro diversity and signal spatial diversity performance they provide. Aiming at the uplink in industrial indoor scenarios, we adopt a path loss model based on real measurement campaigns. Monte Carlo simulation results show that the grid deployment of APs provide higher average channel gains, but radio stripes deployments provide lower variability of the received signal strength.
\end{abstract}


\section{Introduction}


\par The Fifth Generation (5G) of wireless communication systems is the first generation providing native support to Machine-Type Communication (MTC) services in addition to the traditional Human-Type Communication (HTC) services. Academia and industry split MTC use cases into two broad categories: i) massive MTC (mMTC), which aims at providing wireless connectivity to a massive number of Machine-Type Devices (MTDs); and ii) Ultra-Reliable Low-Latency Communications (URLLC), also known as critical MTC (cMTC), which encompasses use cases with very stringent requirements in terms of latency and reliability \cite{METIS5G2016}. However, some of the more demanding requirements of mMTC and cMTC cannot be fully supported by current 5G networks \cite{6G_White_Paper}. 


\par 5G introduced massive Multiple-Input Multiple-Output (mMIMO) technology to enhance the spectral efficiency for HTC use cases, thus enabling the provisioning of very high data rates \cite{interdonato2019}. mMIMO relies in the use of Base Stations (BSs) with a very high number of receive/transmit antenna elements, thus providing great spatial multiplexing capabilities. Recent works have also studied the performance gains that mMIMO can provide to mMTC \cite{liu2018} and cMTC use cases \cite{popovski2019}, \cite{ding2021}. However, as networks become denser, the inter-cell interference, inherent to the traditional cellular paradigm, represents a major bottleneck for the system performance. A promising solution to solve this issue is distributed mMIMO, also knowm as Cell-Free (CF) mMIMO, which consists on the combination of network densification, mMIMO and a novel user-centric design paradigm \cite{interdonato2019}. In a distributed mMIMO system, there are no cell boundaries, thus all the Access Points (APs) in the network jointly cooperate to serve all the users. By using fronthaul connections, the APs are connected to Central Processing Units (CPUs), which perform coordination and signal processing tasks.

\par In \cite{lancho2021}, authors showed that a distributed mMIMO setup with fully centralized processing outperforms the traditional mMIMO setup, in both uplink and downlink, in terms of network availability, which is the probability (computed with respect to the random users' positions) that the error probability is below a given target. 
As also discussed in \cite{lancho2021}, given a total number of antenna elements, it is more beneficial to deploy many single-antenna APs rather than fewer multiple-antenna APs. However, this approach requires a higher fronthaul capacity and also incurs higher deployment costs. In \cite{ma2021}, authors compare the performance of centralized, distributed and cooperative signal processing schemes of a distributed mMIMO network implemented with radio stripes.
Meanwhile, authors in \cite{ganesan2020} compare the performance of centralized mMIMO and distributed mMIMO deployments using computer simulations. They focus on HTC applications in indoor scenarios that require very high data rates, thus considering only the Line of Sight (LOS) case. Their results show that distributed mMIMO setups require less transmit power in the downlink when compared to a centralized mMIMO setup to provide high data rates to several users simultaneously. Using real measurement data obtained in two different factory scenarios, Arnold et al. \cite{arnold2021} demonstrated that a distributed mMIMO setup provides a better signal coverage than a centralized mMIMO setup.

\begin{figure*}[t]
    \centering
    \begin{subfigure}[b]{0.3\textwidth}
        \centering
        \includegraphics[scale=0.4]{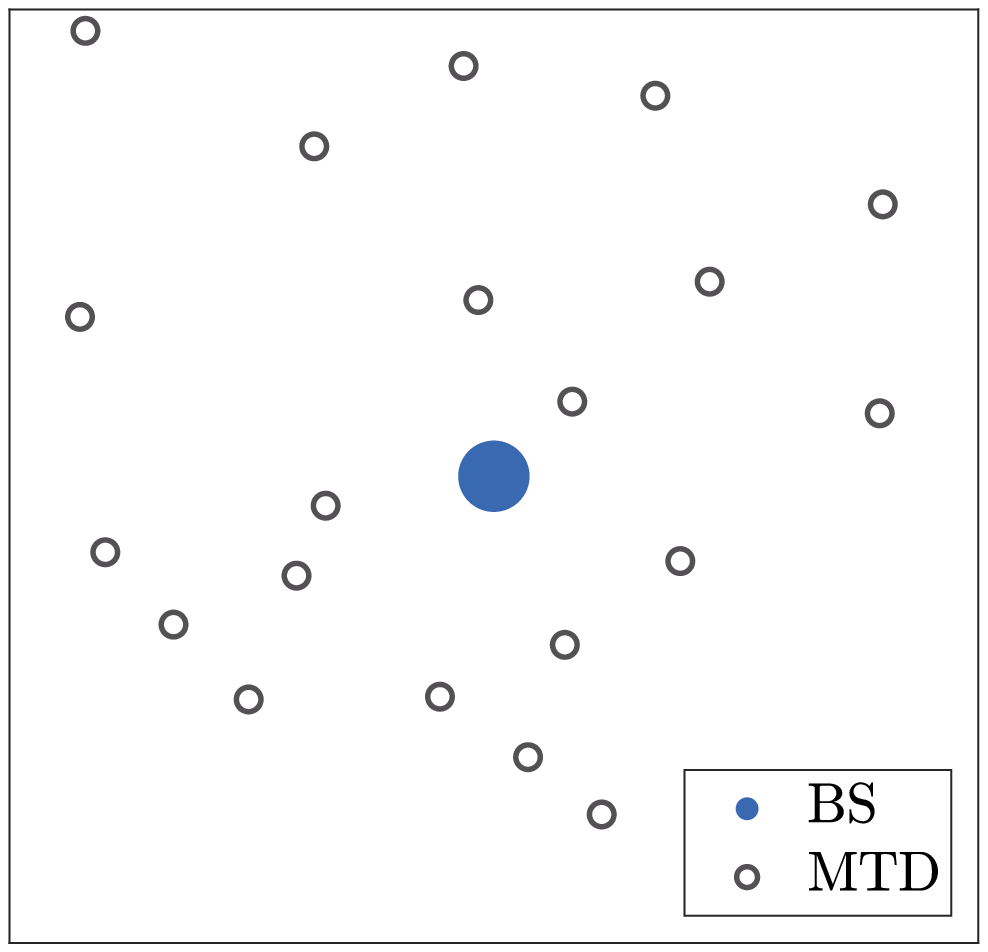}
        \caption{}
        \label{Centralized mMIMO}
    \end{subfigure}
    \hfill
    \begin{subfigure}[b]{0.3\textwidth}
        \centering
        \includegraphics[scale=0.4]{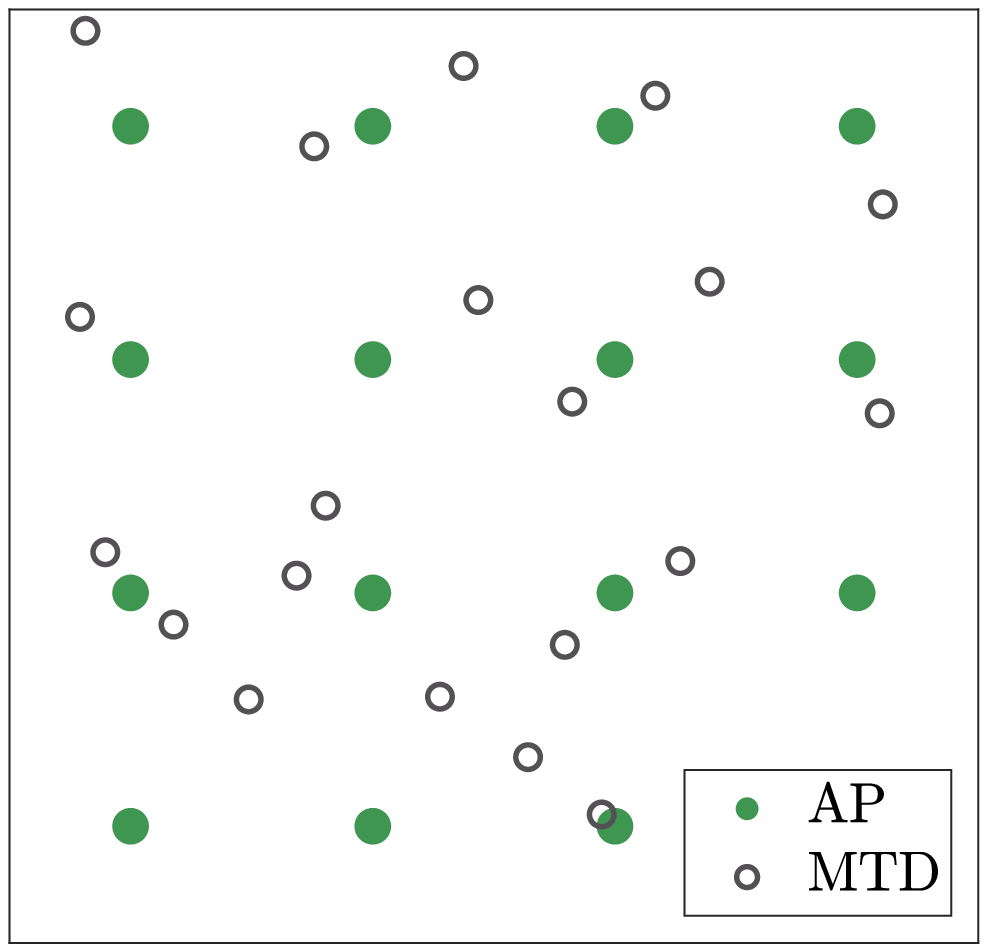}
        \caption{}
        \label{Distributed mMIMO}    
    \end{subfigure}
    \hfill
    \begin{subfigure}[b]{0.3\textwidth}
        \centering
        \includegraphics[scale=0.4]{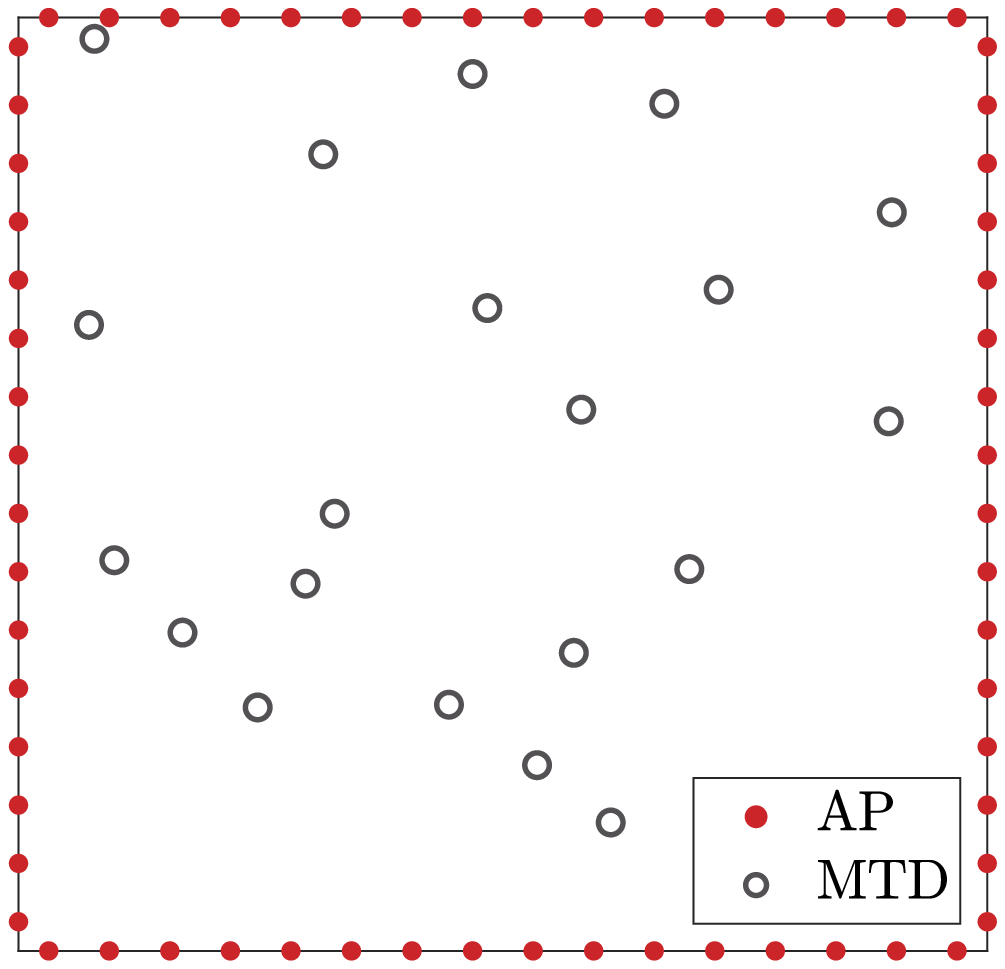}
        \caption{}
        \label{Radio Stripes}
    \end{subfigure}
    \caption{Centralized mMIMO (a), Distributed mMIMO (b) and radio stripes (c) deployments in an indoor square area with sides of length $d$ m. $K$ MTDs are randomly distributed in this area and served by a common BS, in the case of centralized mMIMO, or by $Q$ APs, in the case of distributed mMIMO or radio stripes setups.}
    \label{mMIMO Setups}
\end{figure*}


\par Related works have utilized distributed mMIMO networks in either grid configuration of APs or a radio stripes deployment\footnote{Radio stripe is a low cost implementation of distributed mMIMO proposed by Ericsson in 2019 \cite{Ericsson2019} and that has been studied in some recent works, e.g. \cite{ma2021},\cite{shaik2020},\cite{onel2021}.}, but none of them compared thoroughly the performance achieved with such different settings. In this paper, we introduce performance metrics to compare centralized mMIMO deployments and distributed mMIMO deployments in an indoor industrial scenario. We utilize the average channel gain to measure the macro-diversity gain of each scheme, and the Coefficient of Variation (CV) of the channel gains to measure the variability of the received signal strength. In the case of distributed mMIMO, we compare the performance achieved by different distributed mMIMO deployment schemes: Totally Distributed (TD) mMIMO, where many single-antenna APs are employed, and Partially Distributed (PD) mMIMO, where there is a fewer number of APs but that are equipped with multiple antennas. Thus, differently from the aforementioned works, we compare the performance achieved by distributed mMIMO when APs are employed in grid or radio stripes configurations in terms of macro-diversity gain and variability of the received signal strength in an uplink scenario. The main goal of this work is to find out which deployment of APs is more suitable for Industrial Internet of Things (IIoT) applications with stringent latency and reliability requirements, thus requiring a more predictable wireless channel. Aiming at such industrial use cases, we utilize a path loss model validated by 3GPP for indoor industrial scenarios \cite{Path_Loss_Models_Industrial}. The performance of all the mMIMO deployment schemes are evaluated for the case where an MTD is at a typical position, and for the case where the MTD is located at the worst case position. Our simulation results show that the distributed mMIMO scheme using the grid deployment enhances significantly the average channel gain of the received signal, but the radio stripes configuration is the one that provides the lower variability of the received signal strength, which is a desired characteristic for scalable cMTC scenarios in 6G networks.

\par This paper is organized as follows. The considered system model and centralized/distributed mMIMO schemes are presented in Section \ref{System Model}. Analytical expressions for the average channel gain of a single device achieved with centralized and distributed mMIMO setups are obtained in Section \ref{Closed_Form_Expressions}. Numerical results comparing the performance of the centralized and distributed mMIMO schemes are discussed in Section \ref{Numerical Results}. Finally, we conclude in Section \ref{Conclusions}.

\vspace{-0.1cm}


\section{System Model}
\label{System Model}

\par We consider the uplink of an indoor industrial scenario, which is a factory hall of dimensions $d_X\times d_Y\times h\;\text{m}^3$. We also consider a system that operates with a carrier frequency $f_c=3.5$ GHz since the 3.4-4.2 GHz band is the main candidate for industrial private networks with bandwidths of up to 100 MHz \cite{Path_Loss_Models_Industrial}. 

\par As illustrated in Fig. \ref{mMIMO Setups}, the single-antenna devices in this indoor industrial scenario are served by a total number of $M$ antenna elements, which can be deployed in the factory hall according to one of the following schemes:

\begin{enumerate}
    \item \textit{Centralized mMIMO (Fig. \ref{Centralized mMIMO}):} a single AP with $M$ antenna elements is located at the center of the factory hall. In this case, the indoor scenario can be seen as a single-cell network served by a single BS \cite{casciano2019}.
    \item \textit{Distributed mMIMO (Fig. \ref{Distributed mMIMO}):} $Q$ APs are uniformly distributed on the ceiling of the factory hall in a grid configuration.
    \item \textit{Radio Stripes (Fig. \ref{Radio Stripes}):} low cost deployment of distributed massive MIMO where the antennas and associated processing units (APUs) are serially located inside the same cable, which also provides synchronization, data transfer and power supply via a shared bus \cite{interdonato2019}. A single radio stripe with $Q$ APs is positioned around the walls of the factory hall\footnote{As shown in \cite{ganesan2020}, given a number of antenna elements, the vertical deployment of antenna elements on the wall, that is, using more than one radio stripe on the same wall, does not yield significant performance gains.}.
\end{enumerate}

\par In both distributed mMIMO and radio stripes deployments, each AP is equipped with $S=M/Q$ antenna elements\footnote{In this work, we choose values of $M$ and $Q$ such that $S$ is an integer.}. We denote the case of $S>1$ as a Partially Distributed (PD) setup, and the case of $S=1$ (that is, single antenna APs) as Totally Distributed (TD) setup.

\subsection{Signal Model}

\par Similar to \cite{Book_CF_mMIMO}, we study the signal model for a single active device in the network. 
The $M\times 1$ received signal vector is
\begin{equation}
    \textbf{y}=\sqrt{P_T}\textbf{g}x+\textbf{n},
\end{equation}
where $P_T$ is the transmit power of the MTD, $\textbf{g}\in\mathbb{C}^{M\times 1}$ is the channel vector between the $M$ antenna elements and the device antenna, $x\in\mathbb{C}$ is the transmitted symbol with $\mathbb{E}\{|x|^2\}=1$, $\textbf{n}\in\mathbb{C}^{M\times 1}$ is the vector of Additive White Gaussian Noise (AWGN) samples, such that $\textbf{n}\sim\mathcal{CN}(\textbf{0}_{M\times1},\sigma_N^2\textbf{I}_M)$, and $\sigma_N^2$ is the variance of the noise.


\par The vector of wireless channel coefficients between the device and the $Q$ APs is
\begin{equation}
    \label{g}
    \textbf{g}=[\textbf{g}_{1}^T,\textbf{g}_{2}^T,\ldots,\textbf{g}_{Q}^T]^T \in \mathbb{C}^{M\times1},
\end{equation}
where
\begin{equation}
    \label{g_q}
    \textbf{g}_{q} = \sqrt{\beta_{q}}\textbf{h}_{q}^T\in \mathbb{C}^{S\times1}
\end{equation}
is the vector of wireless channel coefficients between the device and the $q$-th AP, $\beta_{q}$ is the large scale fading coefficient between the device and the $q$-th AP, and $\textbf{h}_{q}\in\mathbb{C}^{S\times1}$ is the vector of small scale fading coefficients from the device to the $q$-th AP. For simplicity, we assume that the CPU has perfect CSI\footnote{In the uplink, the CSI is acquired by the use of pilot sequences transmitted by the devices that are known by the CPU.}.

\subsection{Large Scale Fading}

\par In order to consider the impact of the large scale fading in the indoor industrial scenario, we adopt the model from \cite{Path_Loss_Models_Industrial}. This model is supported by a measurement campaign at $3.5$ GHz in two different types of industrial scenarios and for different APs and devices deployments. As in \cite{casciano2019}, we consider the ``dense factory clutter" scenario with ``clutter embedded APs", which characterizes the most harsh channel conditions. This setup assumes that there is No Line-of-Sight (NLOS) between the APs and served devices. Due to this fact and also to the high abundance of industrial equipment with reflective metallic materials in the factory hall, we assume a Rayleigh small scale fading model, thus $\textbf{g}_{q}\sim\mathcal{CN}(\textbf{0}_{1\times S},\beta_{q}\textbf{I}_S)$, where $\beta_{q}<1$ is the large scale fading term that accounts for path loss and shadowing.

\par In the considered scenario, the path loss (in dB) is modeled as \cite{Path_Loss_Models_Industrial}
\begin{equation}
    \label{path loss}
    \overline{\text{PL}}_{\text{dB}}=32.5+20\log_{10} f_c +10 \eta\log_{10} d_{\text{3D}} \text{ [dB]},
\end{equation}
where $f_c$ is the carrier frequency in GHz, $\eta=3.19$ is the path loss exponent, and $d_{\text{3D}}$ is the 3D distance between the AP and the MTD in meters.

\par The total attenuation due to distance and shadowing is then given by
\begin{equation}
    \label{PL_dB}
    \text{PL}_{\text{dB}} = \overline{\text{PL}}_{\text{dB}} + X_{\text{dB}}\text{ [dB]},
\end{equation}
where $X_{\text{dB}}\sim\mathcal{N}(0,\sigma_S^2)$ is a log-normal random variable that represents the shadowing term with $\sigma_S=7.56$ dB \cite{Path_Loss_Models_Industrial}.

\par The linear model for the large scale fading coefficient between the device and the $q$-th AP is finally given by
\begin{equation}
    \label{beta}
    \beta_{q}(d_{q},\sigma_S)=\dfrac{1}{\text{PL}}=\dfrac{10^{X_{\text{dB}}/10}}{\overline{\text{PL}}(d_{q})},
\end{equation}
where PL is the total attenuation due to path loss and shadowing in linear scale, and $\overline{\text{PL}}$ is the path loss in linear scale.

\section{Macro Diversity Gain Analysis}
\label{Closed_Form_Expressions}

\par In this section, we present closed-form expressions of the average channel gain $\mathbb{E}\left\{\lVert\textbf{g}\rVert\right\}^2$ for the considered mMIMO setups. The objective of this analysis is to quantify the macro diversity gain obtained with distributed mMIMO when compared to centralized mMIMO. Let us define $Z\triangleq10^{X_{\text{dB}}/10}$, which is the shadowing in linear scale. Then, the expected value of the shadowing in linear scale can be calculated as
\begin{align}
    \mathbb{E}\left\{Z\right\}&=\int_{-\infty}^{\infty} 10^{Z/10} \dfrac{1}{\sqrt{2\pi\sigma_S^2}}\exp\left(-\dfrac{Z^2}{2\sigma_S^2}\right)dZ\notag\\
    &= \dfrac{1}{\sqrt{2\pi\sigma_s^2}}\exp\left(\dfrac{2\sigma_s^2[\ln(10)]^2}{400}\right) \times \notag\\ & \int_{-\infty}^{\infty}\exp\left\{-\left(\dfrac{Z}{2\sigma_s^2}-\dfrac{\sqrt{2\sigma_s^2}}{20}\ln(10)\right)^2 \right\}dZ\notag\\
    &= \exp\left(\dfrac{2\sigma_s^2[\ln(10)]^2}{400}\right) \times \notag\\
    & \int_{-\infty}^{\infty}\dfrac{1}{\sqrt{2\pi\sigma_s^2}}\exp\left(-\dfrac{[Z-2\sigma_s^2\ln(10)/20]^2}{2\sigma_s^2} \right)dZ \notag\\
    &=10^{2\sigma_S^2\log(10)/400},
\end{align}
where we rewrote the integrand to include the Gaussian Probability Density Function (PDF), which integrates to 1, and performed simple algebraic transformations.

\par Finally, using the transformed random variable in (\ref{beta}), the expected value of the large-scale fading coefficient between an MTD and an AP antenna element is
\begin{equation}
    \label{expected_value_beta}
    \mathbb{E}\left\{\beta_{q}\right\}=\dfrac{\mathbb{E}\left\{Z\right\}}{\overline{\text{PL}}(d_{q})}
    =10^{2\sigma_S^2\log(10)/400}\dfrac{1}{\overline{\text{PL}}(d_{q})}.
\end{equation}

\subsection{Centralized mMIMO}

\par In the centralized mMIMO setup, all the antenna elements are co-located at the same position, thus the large scale fading coefficient between the MTD and each one of the $M$ antenna elements is approximately the same. Thus, we assume $\beta_q=\beta\;\forall q$. The channel vector can be written as $\textbf{g}=\beta\lVert\textbf{h}\rVert^2$. Using (\ref{expected_value_beta}), and the fact that $\mathbb{E}\left\{\lVert\textbf{h}_k\rVert^2 \right\}=M$ is the array gain owing to the $M$ antennas at the BS, the average channel gain is given by
\begin{align}
    \label{average_channel_gain_centralized_mMIMO}
    \mathbb{E}\left\{\lVert\textbf{g}\rVert^2\right\}&=\mathbb{E}\left\{\beta\lVert\textbf{h}\rVert^2 \right\} = \mathbb{E}\left\{\beta\right\} \cdot \mathbb{E}\left\{\lVert\textbf{h}\rVert^2 \right\}\notag\\
    &=10^{2\sigma_S^2\log(10)/400}\dfrac{M}{\overline{\text{PL}}(d_{\text{3D}})}.
\end{align}    

\subsection{Distributed mMIMO}

\par We now derive an expression for any distributed mMIMO setup. The channel vector between an MTD and the $Q$ APs was given by (\ref{g}). Then, the average channel gain is
\begin{align}
    \label{average_channel_gain_distributed_mMIMO}
    \mathbb{E}\left\{\lVert\textbf{g}\rVert^2\right\}&=\sum_{q=1}^Q \mathbb{E}\left\{\lVert\textbf{g}_{q}\rVert^2\right\}=
    \sum_{q=1}^Q  \mathbb{E}\left\{\beta_{q} \right\}\mathbb{E}\left\{\lVert\textbf{h}_{q}\rVert^2\right\}\notag\\
    &=10^{2\sigma_S^2\log(10)/400} \sum_{q=1}^{Q} \dfrac{S}{\overline{\text{PL}}(d_{q})}.
\end{align}

\par The expected value of each large-scale fading coefficient $\mathbb{E}\left\{\beta_{q} \right\}$ is computed using (\ref{expected_value_beta}). We also used the fact that $\mathbb{E}\left\{\lVert\textbf{h}_{q}\rVert^2 \right\}=S$ is the array gain owing to the $S$ antenna elements in each AP. The summation in (\ref{average_channel_gain_distributed_mMIMO}) corresponds to the macro-diversity gain owing to the spatial distribution of the antenna elements.

\begin{table*}[t!]
    \centering
    \caption{Expected value and standard deviation of the channel gain for an MTD at a typical position in the factory hall and for all the mMIMO deployments.}
    \label{Table_Typical_Case}
    \begin{tabular}{|c|c|c|c|c|c|}
        \hline
        \textbf{Parameter} & \makecell{\textbf{Centralized}\\\textbf{mMIMO}} & \makecell{\textbf{PD}\\\textbf{mMIMO}} & \makecell{\textbf{TD}\\\textbf{mMIMO}} & \makecell{\textbf{PD}\\\textbf{Radio Stripes}} & \makecell{\textbf{TD}\\\textbf{Radio Stripes}} \\
        \hline
        $\mathbb{E}\left\{\lVert\textbf{g}\rVert^2\right\}$ & $-63.6114$ dB & $-63.7807$ dB & $-60.3654$ dB & $-71.5584$ dB & $-71.5475$ dB\\
        \hline
        $\sigma_g$ & $-57.6406$ dB & $-60.9926$ dB & $-56.6184$ dB & $-68.7296$ dB & $-70.4862$ dB \\
        \hline
        $c_v=\sigma_g/\mathbb{E}\left\{\lVert\textbf{g}\rVert^2\right\}$ & 3.9544 & 1.9002 & 2.3697 & 1.9181 & 1.2768 \\
         \hline
    \end{tabular}    
\end{table*}

\begin{table*}[t!]
    \centering
    \caption{Expected value and standard deviation of the channel gain for an MTD at the worst case position in the factory hall and for all the mMIMO deployments.}
    \label{Table_Worst_Case}
    \begin{tabular}{|c|c|c|c|c|c|}
        \hline
        \textbf{Parameter} & \makecell{\textbf{Centralized}\\\textbf{mMIMO}} & \makecell{\textbf{PD}\\\textbf{mMIMO}} & \makecell{\textbf{TD}\\\textbf{mMIMO}} & \makecell{\textbf{PD}\\\textbf{Radio Stripes}} & \makecell{\textbf{TD}\\\textbf{Radio Stripes}} \\
        \hline
        $\mathbb{E}\left\{\lVert\textbf{g}\rVert^2\right\}$ & $-77.5595$ dB & $-69.8238$ dB & $-67.2229$ dB & $-74.4385$ dB & $-74.4755$ dB\\
        \hline
        $\sigma_g$ & $-71.5887$ dB & $-64.4698$ dB & $-62.3856$ dB & $-74.0119$ dB & $-75.4007$ dB \\
        \hline
        $c_v=\sigma_g/\mathbb{E}\left\{\lVert\textbf{g}\rVert^2\right\}$ & 3.9544 & 3.2999 & 3.0460 & 1.1032 & 0.8081\\
         \hline
    \end{tabular}    
\end{table*}

\par Comparing (\ref{average_channel_gain_centralized_mMIMO}) and (\ref{average_channel_gain_distributed_mMIMO}), we observe that the distributed mMIMO setup exchanges the array gain of a BS with many antenna elements by the macro diversity gain owing to the spatial distribution of many APs.

\section{Numerical Results}
\label{Numerical Results}

\begin{figure}[t!]
    \centering
    \subfloat[]{\label{Results_Channel_Gains_Typical_Case}\includegraphics[scale=0.5]{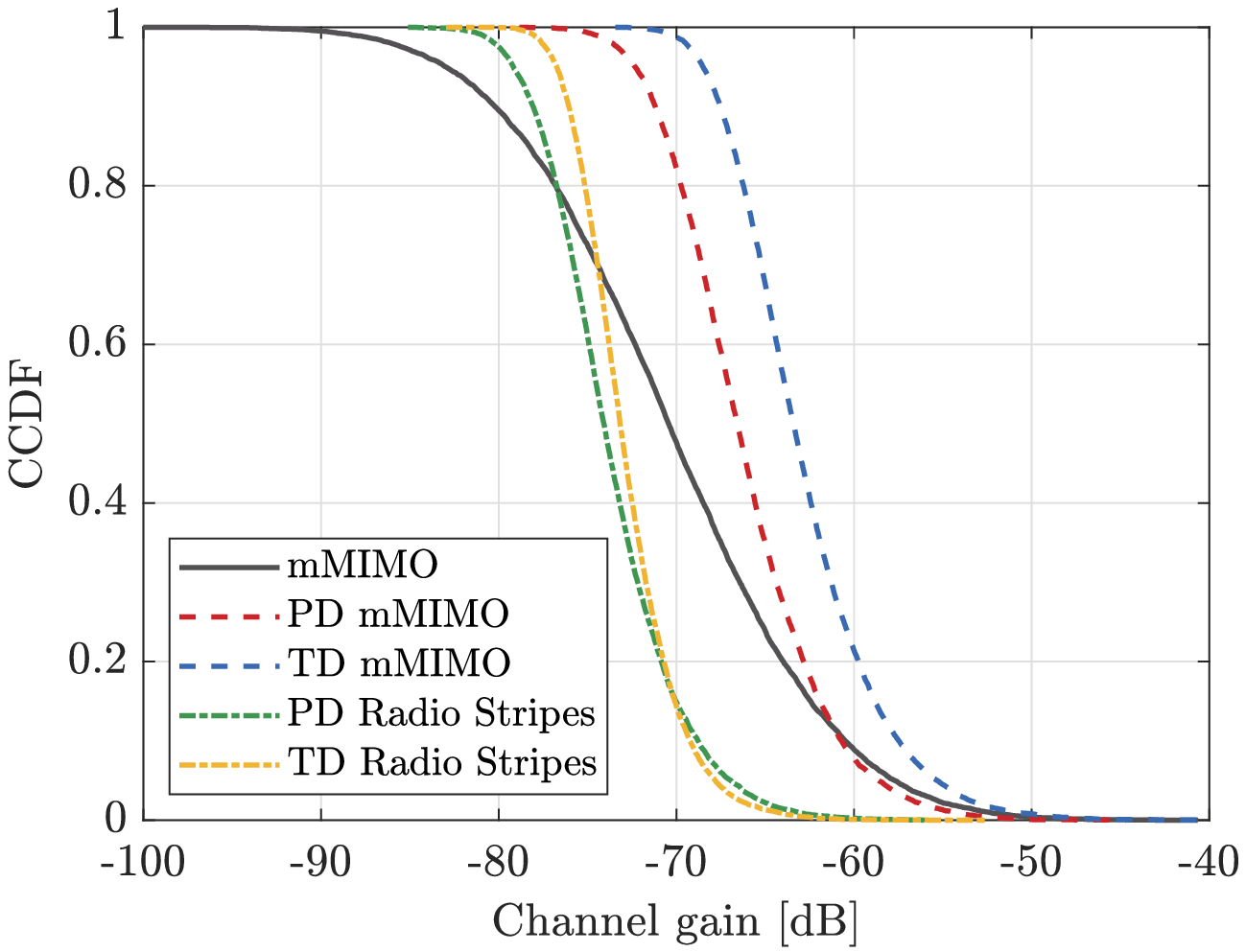}} \\
    \subfloat[]{\label{Results_Channel_Gains_Worst_Case}\includegraphics[scale=0.5]{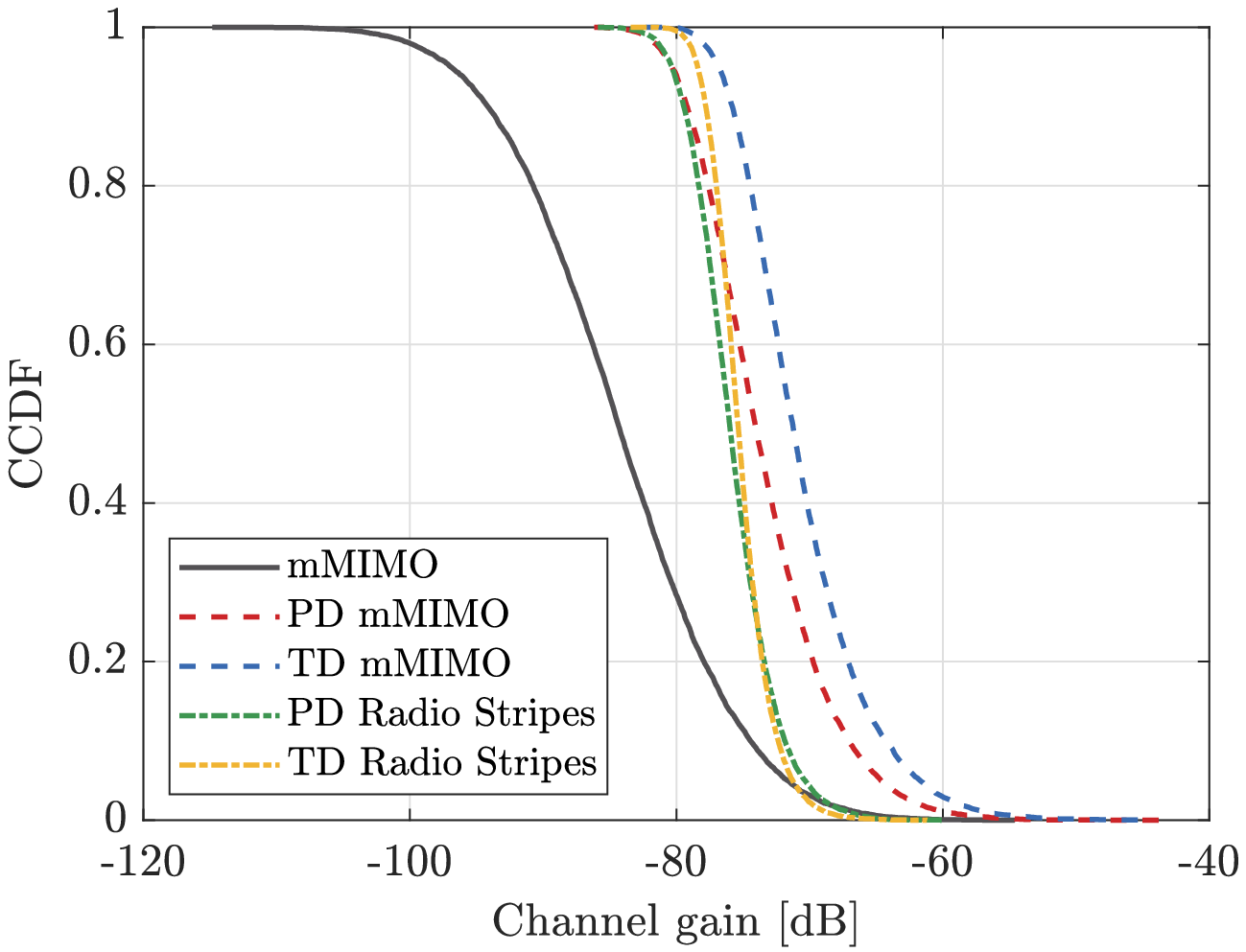} }
    \caption{Empirical CCDF of the channel gains for the MTD at a (a) typical position  and (b) worst case position in the factory hall and for all the different mMIMO deployments.}
    \label{Results_Channel_Gains}
\end{figure}

\par In this section, we compare the performance achieved under the different mMIMO settings presented in Section \ref{System Model} using Monte Carlo simulations.  Similar to \cite{casciano2019} and \cite{ding2021}, we consider a square indoor industrial scenario with sides of length $d=100$ m and two different situations: i) the MTD is located at a typical position $d_{\text{MTD}}=[55\text{ m, }75\text{ m}]$, and ii) the worst case position, which depends on the specific mMIMO deployment. In the case of centralized mMIMO or mMIMO using the grid deployment, the worst position for the MTD is any of the corners of the square area. On the other hand, in the case of radio stripes deployment, the worst position is the center of the square area. 

\par 
As the simulation parameters, we set $f_c=3.5$ GHz and $M=64$.
In the centralized mMIMO deployment, the BS is located at the position $d_{\text{BS}}=[50\text{ m, }50\text{ m}]$. For the grid deployments, we evaluate the performance of a PD mMIMO deployment with $Q=16$ APs, each equipped with $S=4$ antennas, and TD mMIMO deployment with $Q=64$ single-antenna APs. Similarly, we also evaluate the performance of a PD radio stripe deployment with $Q=16$ APs, each equipped with $S=4$ antennas, and a TD radio stripe deployment with $Q=64$ single-antenna APs. In all cases, the height of the APs is $h_{\text{AP}}=6$ m, and the MTD is at a height $h_{\text{MTD}}=1.5$ m.

\subsection{Macro Diversity Gain Analysis}

\begin{table*}[]
    \centering
    \caption{Expected value, standard deviation and ratio of the channel gains for an MTD at a typical position in the factory hall and for the TD mMIMO setup with $M=64$ and $S=1$.}
    \label{Table_Typical_Case_Subset_APs}
    \begin{tabular}{|c|c|c|c|c|c|}
        \hline
        \textbf{Parameter} & $\mathbf{|Q_{AP}|=1}$ & $\mathbf{|Q_{AP}|=4}$ & $\mathbf{|Q_{AP}|=8}$ & $\mathbf{|Q_{AP}|=16}$ & \textbf{All APs} \\
        \hline
        $\mathbb{E}\left\{\lVert\textbf{g}\rVert^2\right\}$ & $-65.1311$ dB & $-61.5440$ dB & $-60.9829$ dB & $-60.6403$ dB & $-60.3654$ dB\\
        \hline
        $\sigma_g$ & $-57.6394$ dB & $-56.6582$ dB & $-56.6305$ dB & $-56.6199$ dB & $-56.6184$ dB \\
        \hline
        $c_v=\sigma_g/\mathbb{E}\left\{\lVert\textbf{g}\rVert^2\right\}$ & 5.6127 & 3.0802 & 2.7242 & 2.5237 & 2.3697\\
        \hline
        \makecell{Ratio Channel Gains} & $0.21297$ & $0.57405$ & $0.73147$ & $0.86281$ & $1$\\
         \hline 
    \end{tabular}    
\end{table*}

\begin{table*}[]
    \centering
    \caption{Expected value, standard deviation and ratio of the channel gains for an MTD at the worst case position in the factory hall and for the TD mMIMO setup with $M=64$ and $S=1$.}
    \label{Table_Worst_Case_Subset_APs}
    \begin{tabular}{|c|c|c|c|c|c|}
        \hline
        \textbf{Parameter} & $\mathbf{|Q_{AP}|=1}$ & $\mathbf{|Q_{AP}|=4}$ & $\mathbf{|Q_{AP}|=8}$ & $\mathbf{|Q_{AP}|=16}$ & \textbf{All APs} \\
        \hline
        $\mathbb{E}\left\{\lVert\textbf{g}\rVert^2\right\}$ & $-68.8524$ dB & $-67.8685$ dB & $-67.5936$ dB & $-67.4158$ dB & $-67.2229$ dB\\
        \hline
        $\sigma_g$ & $-62.4535$ dB & $-62.3929$ dB & $-62.3870$ dB & $-62.3821$ dB & $-62.3856$ dB \\
        \hline
        $c_v=\sigma_g/\mathbb{E}\left\{\lVert\textbf{g}\rVert^2\right\}$ & 4.3640 & 3.5283 & 3.3164 & 3.1840 & 3.0460\\
        \hline
        \makecell{Ratio Channel Gains} & $0.38976$ & $0.65029$ & $0.77207$ & $0.86981$ & $1$\\
         \hline 
    \end{tabular}    
\end{table*}

\begin{figure}
    \centering
    \subfloat[]{\label{Results_Subset_APs_Typical}\includegraphics[scale=0.5]{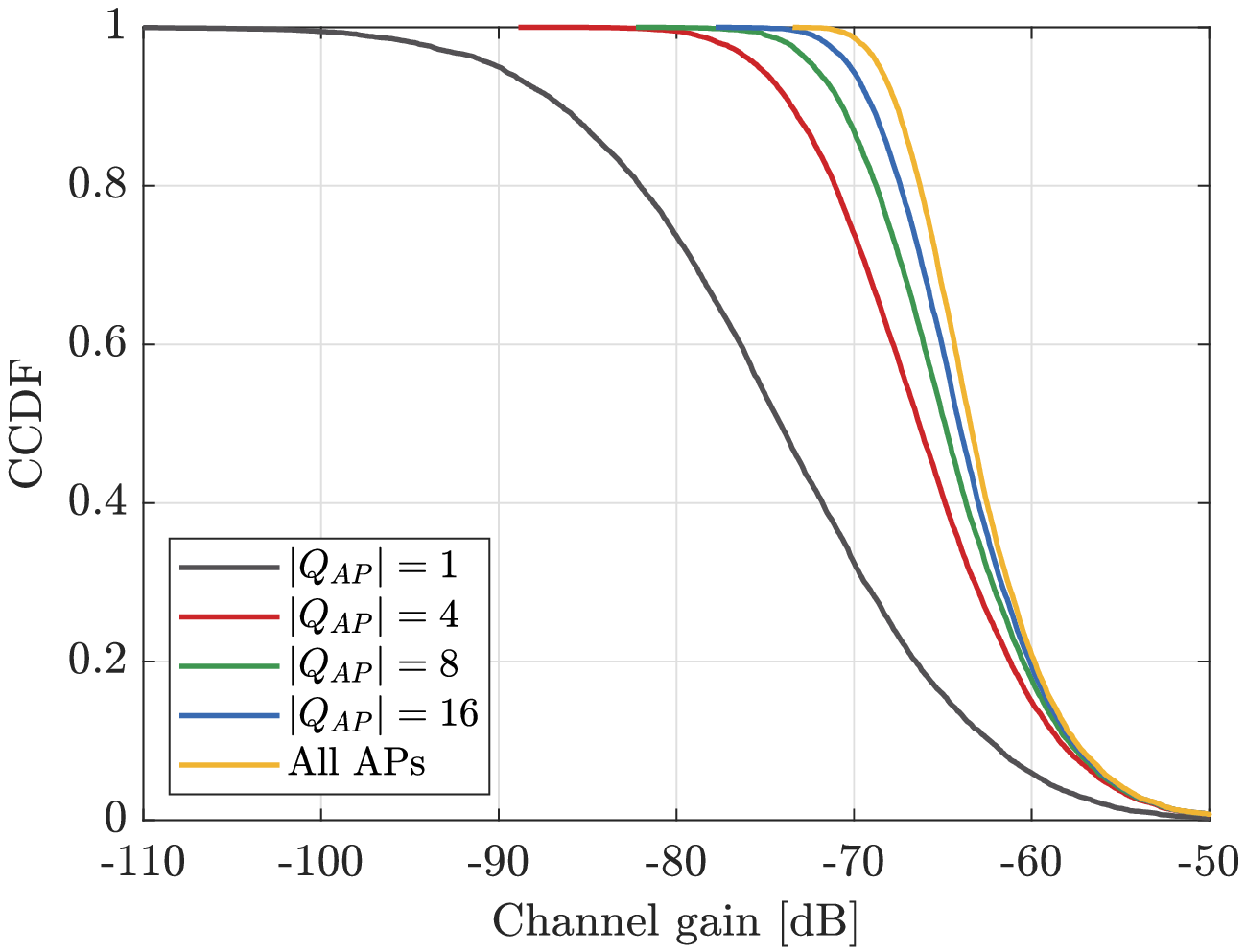}} \\
    \subfloat[]{\label{Resuls_Subset_APs_Worst_Case}\includegraphics[scale=0.5]{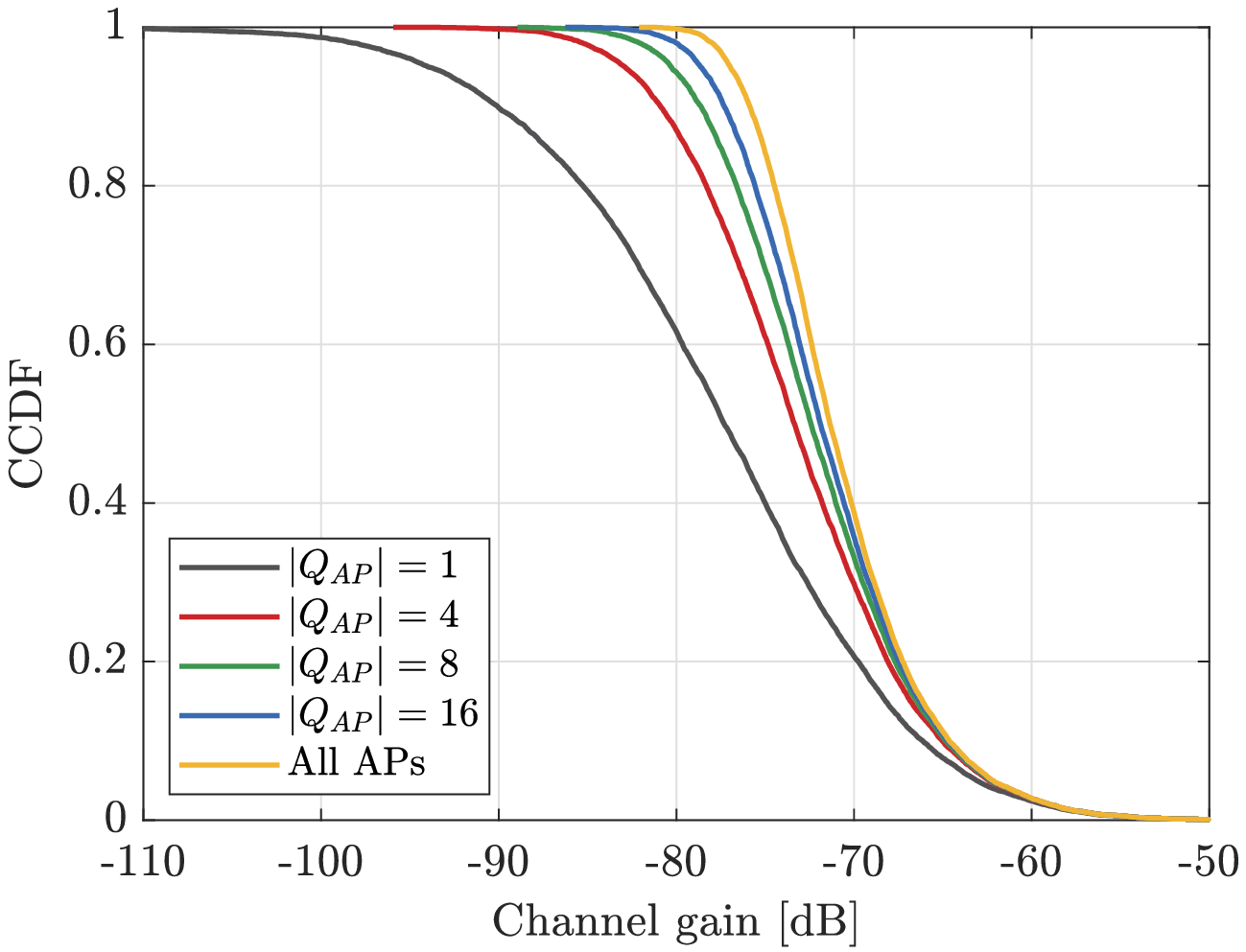} }
    \caption{Empirical CCDF of the channel gains of an MTD at a (a) typical position and (b) the worst case position in the factory hall for the distributed mMIMO setting and $M=64$, $S=1$.}
    \label{Resuls_Subset_APs}
\end{figure}

\par Fig. \ref{Results_Channel_Gains_Typical_Case} shows the empirical Complementary Cumulative Distribution Function (CCDF) of the channel gains for the typical case, while Fig. \ref{Results_Channel_Gains_Worst_Case} shows the same results for the worst case. Moreover, Tables \ref{Table_Typical_Case} and \ref{Table_Worst_Case} list the expected value\footnote{The expected values of the average channel gains obtained in our simulations match the theoretical values obtained using the closed form expressions derived in Section \ref{Closed_Form_Expressions}.}, standard deviation $\sigma_g$ and CV of the channel gains for both cases. The CV, also known as relative standard deviation, is a normalized measure of of the dispersion of a PDF and is given by
\begin{equation}
    c_v=\dfrac{\sigma_g}{\mathbb{E}\left\{\lVert\textbf{g}\rVert^2\right\}}.  
\end{equation}
The CV is a metric that measures the variability of the received signal strength. Less variability can be exploited for more efficient utilization of spectrum resources.

\par When we analyze the results of the typical case in Fig. \ref{Results_Channel_Gains_Typical_Case} and in Table \ref{Table_Typical_Case}, we note that the radio stripes configurations achieve the lowest average channel gains, while the distributed mMIMO setups with grid distribution achieve the highest average channel gain. However, the radio stripe settings present the lowest standard deviation of the instantaneous channel gains, which means that the received signal strength presents less fluctuations. The centralized mMIMO setup achieves an average channel gain that is better than the setups with radio stripes and worse than the grid setups, but it presents the highest standard deviation, which means that the received signal strength is subject to a high variability. Nevertheless, as can be seen in Table \ref{Table_Typical_Case}, the lowest value of CV is achieved by the PD mMIMO setup because the MTD is very close to a multi-antenna AP in this case.

\par We now consider the results for the worst case scenario shown in Fig. \ref{Results_Channel_Gains_Worst_Case} and Table \ref{Table_Worst_Case}. We observe that the lowest average channel gain and the higher standard deviation are achieved by the centralized mMIMO scheme. The grid configuration still outperforms the radio stripes configurations for both PD and TD mMIMO setups in terms of average channel gain, but now the radio stripes configurations also outperform the centralized mMIMO setup. As shown by the CV values in Table \ref{Table_Worst_Case}, the PD and TD radio stripe deployments provide the received signal strength with lower variability.

\par By analyzing the results of the distributed mMIMO schemes using both grid or radio stripes setups, we conclude that the more the antenna elements are distributed, the higher is the average channel gain and the lower is its variability. This happens because the MTD is likely to be very close to one AP or to subset of APs, thus the average distance from the MTD to any nearest AP is much shorter compared to the distance between the MTD and the BS in the centralized mMIMO. When we utilize radio stripes, the APs are only located on the walls of the factory hall, thus the average distance between the MTD and the APs is higher when compared to the distributed mMIMO scheme where the APs are located on the ceiling. On the other hand, the installation and maintenance costs of the radio stripe system is lower when compared to the traditional distributed mMIMO, since the latter requires much more fronthaul connections between the APs and the CPU\footnote{When sequential uplink processing is adopted, only a single front-haul connection between the radio stripe and the CPU is required \cite{shaik2020}.}. 
The grid configuration requires higher amount of signaling between the APs and the CPU. The PD mMIMO schemes presents an slightly inferior performance, but they can significantly reduce the required amount of fronthaul signaling when compared to TD deployments. This fact evinces the existence a trade-off between the achieved macro-diversity gain and the required fronthaul capacity.

\subsection{Signal Spatial Diversity}

\par In this subsection, we consider only the TD mMIMO setting with $Q=64$ single-antenna APs. We assume that the CPU has perfect knowledge of the large scale fading coefficients between the MTD and all the APs, thus it can select the subset of closest APs to decode its signal. Let $\mathcal{Q}_{AP}$ denote this subset, and let $|\mathcal{Q}_{AP}|$ denote its cardinality. In Figs. \ref{Results_Subset_APs_Typical} and \ref{Resuls_Subset_APs_Worst_Case}, we plot the empirical CCDF of the channel gains for the typical and worst cases, respectively, and for $|\mathcal{Q}_{AP}|\in\left\{1,4,8,16\right\}$\footnote{The case where $|\mathcal{Q}_{AP}|=1$ can be interpreted as a small cells deployment: the MTD is connected to the closest AP, which behaves as a BS covering a small square cell of dimensions $d/\sqrt{Q} \times d/\sqrt{Q}$.}. Moreover, Tables \ref{Table_Typical_Case_Subset_APs} and \ref{Table_Worst_Case_Subset_APs} list the average channel gain, the standard deviation of the channel gains, and the CV values for the typical and worst cases, respectively. These tables also show the mean value of the ratio between the channel gain achieved when only the APs in $Q_{AP}$ are used to decode the signal from the MTD and the channel gain achieved when all the APs are used. Thus, this ratio is a measure of how much energy of the signal transmitted by the MTD is captured by the APs in $Q_{AP}$.

\par We observe that the greater the number of APs is used to decode the signal from the MTD, the greater the average channel gain and the lower the variability of the received signal strength are. However, the performance gains are much more accentuated in the typical case when compared to the worst case. In the latter case, the performance achieved with $|Q_{AP}|=16$  is very close to the performance achieved when all APs are used. Nevertheless, utilizing only a subset of APs to serve an MTD reduces significantly the required fronthaul capacity for signal processing.

\par Note also that the subset of neighboring APs captures a significant portion of the energy  of the signal transmitted by the MTD. In other words, only the neighboring APs within the communication range of the MTD have non-negligible channel gains due to the macro-diversity, which leads to the signal spatial diversity in distributed mMIMO. By using this fact, if the appropriate subset of APs is chosen to decode the signal from an MTD, and if the subsets of APs for different MTDs are different or only partially overlapped, the interference in a multi-user setup be significantly reduced when compared to the centralized mMIMO deployment \cite{casciano2019}.



\section{Conclusions}
\label{Conclusions}

\par In this paper, we compared the performance of centralized mMIMO and different distributed mMIMO setups for the typical and worst cases of each scheme. Our simulation results showed that distributed mMIMO schemes using grid deployments provide a highest average channel gain, which is interesting for applications that demand high data rates, but distributed mMIMO using the radio stripes deployment provides a lower variability of the received signal strength, which may be desirable for applications with more stringent latency and reliability requirements such as cMTC in beyond-6G and 6G networks. We showed that the more distributed the antenna elements are, the higher the achieved macro-diversity gain is. 
Finally, we also illustrated the signal spatial diversity achieved by the distributed mMIMO using the grid configuration. By using only the subset of closest APs to decode the signal transmitted by a single MTD, we still achieve a satisfactory performance while at the same time reducing significantly the required fronthaul capacity. Moreover, the proper choice of APs used to the decode the signal transmitted by an MTD opens the path for robust interference mitigation mechanisms based on the spatial signal diversity provided by distributed mMIMO. The framework of this paper is scalable, that is, even increasing the physical dimensions, number of antenna elements or number of devices, the relative performance among the schemes shall be the same.

\section*{Acknowledgment}

\par This research has been financially supported by Academy of Finland, 6Genesis Flagship (grant no. 318937), European Union’s Horizon 2020 research and innovation programme (EU-H2020), Hexa-X project (grant no. 101015956) and CNPq (Brazil).

\bibliographystyle{./bibliography/IEEEtran}
\bibliography{./bibliography/references}

\end{document}